\documentclass[a4paper,fleqn,usenatbib]{mnras}

\usepackage[T1]{fontenc}
\usepackage{ae,aecompl}

\usepackage{graphicx}	% Including figure files
\usepackage{amsmath}	% Advanced maths commands
\usepackage{amssymb}	% Extra maths symbols

\title[Lithium]{Evolution of lithium in the Milky Way halo, discs and bulge}

\author[Grisoni et al.]{V. Grisoni$^{1,2}$\thanks{E-mail: valeria.grisoni@inaf.it}, F. Matteucci$^{1, 2, 3}$, D. Romano$^4$, X. Fu$^5$\\
 $^1$ Dipartimento di Fisica, Sezione di Astronomia, Universit\`a di Trieste, via G.B. Tiepolo 11, I-34131, Trieste, Italy \\  
 $^2$ INAF Osservatorio
  Astronomico di Trieste, via G.B. Tiepolo 11, I-34131, Trieste,
  Italy\\
 $^3$ INFN Sezione di Trieste, via Valerio 2, 34134 Trieste, Italy\\
 $^4$ INAF Osservatorio di Astrofisica e Scienza dello Spazio, Via Gobetti 93/3, I-40129 Bologna, Italy\\
 $^5$ The Kavli Institute for Astronomy and Astrophysics at Peking University, Beijing, China
}

\begin{document}
\date{Accepted . ; in original form xxxx}

\pagerange{\pageref{firstpage}--\pageref{lastpage}} \pubyear{xxxx}

\maketitle

\label{firstpage}

\begin{abstract}
In this work, we study the Galactic evolution of lithium by means of chemical evolution models in the light of the most recent spectroscopic data from Galactic stellar surveys. We consider detailed chemical evolution models for the Milky Way halo, discs and bulge, and we compare our model predictions with the most recent spectroscopic data for these different Galactic components. In particular, we focus on the decrease of lithium at high metallicity observed by the AMBRE Project, the Gaia-ESO Survey, and other spectroscopic surveys, which still remains unexplained by theoretical models. We analyse the various lithium producers and confirm that novae are the main source of lithium in the Galaxy, in agreement with other previous studies. Moreover, we show that, by assuming that the fraction of binary systems giving rise to novae is lower at higher metallicity, we can suggest a novel explanation to the lithium decline at super-solar metallicities: the above assumption is based on independent constraints on the nova system birthrate, that have been recently proposed in the literature. As regards to the thick disc, it is less lithium enhanced due to the shorter timescale of formation and higher star formation efficiency with respect to the thin disc and, therefore, we have a faster evolution and the "reverse knee" in the A(Li) vs. [Fe/H] relation is shifted towards higher metallicities. Finally, we present our predictions about lithium evolution in the Galactic bulge, that, however, still need further data to be confirmed or disproved.
\end{abstract}


\section{Introduction}

We are in a golden era for Galactic Archaeology thanks to the advent of large spectroscopic surveys from the ground and space missions. A huge amount of data is being collected, which is boosting the number of open questions that have to be solved by theoretical models.
\\Among these, one of the most puzzling topics is the Galactic lithium evolution. Understanding the evolution of this element, indeed, raises a number of difficult questions, as briefly recalled in the following paragraphs.
\\Starting from the pioneering work of Spite \& Spite (1982), it has been generally acknowledged  that most metal-poor (-2.4 < [Fe/H] < -1.4), warm (T$_{eff}$ = 5700-6800 K) Galactic halo dwarfs lie on a well-defined plateau, the so-called "Spite plateau", namely, they share roughly the same Li abundance, A(Li)=2.05-2.2 dex (Spite \& Spite 1986; Bonifacio \& Molaro 1997; Bonifacio et al. 2007). This common abundance, however, is i) lower than that predicted by standard Big Bang neocleosynthesis theory (SBBN) as inidicated by the baryon density (see e.g. the Planck results, Coc et al. 2014) and ii) lower than that observed in meteorites (e.g. Lodders et al. 2009) and young T Tauri stars (Bonsack \& Greenstein 1960). To make the story a bit more complicated, a huge dispersion in Li abundances is observed for stars at disc metallicities (see Ramirez et al. 2012; Delgado Mena et al. 2015; Guiglion et al. 2016; Fu et al. 2018, among others, for recent work), while observations of extremely metal-poor stars (Sbordone et al. 2010; Melendez et al. 2010; Hansen et al. 2014; Bonifacio et al. 2015) find the Spite plateau to bend down for stars with [Fe/H]<-2.8 dex. In this work, we do not consider the first problem (the so-called cosmological Li problem). This is quite convincingly addressed in Fu et al. (2015), which explain the discrepancy between the SBBN Li value and the one observed on the plateau, as well as the drop of Li abundances at very low metallicities, as due to stellar mechanisms. We rather concentrate on the second problem, i.e. the Galactic evolution of lithium. As customarily done in the literature, we assume that the upper envelope of the observations in a A(Li)-[Fe/H] diagram faithfully traces the enrichment of Li in time in the Galaxy.
\\A topic of lively debate is the interstellar medium (ISM) lithium content decline observed at super-solar metallicities for solar neighbourhood stars. This feature was first pointed out by Delgado Mena et al. (2015), and then confirmed by the AMBRE Project (Guiglion et al. 2016), the Gaia-ESO Survey -hereinafter GES- (Fu et al. 2018), Bensby \& Lind (2018), and can be also seen in the recent GALAH DR2 data (Buder et al 2018). The scenarios proposed to explain this feature from the Galactic Chemical Evolution (GCE) point of view can be summarized as follows: i) lower yields of Li from stars at high metallicities, even if no physical reasons for this fact can be found (Prantzos et al. 2017); and ii) the interplay of different populations coming from the inner regions of the Milky Way disc (Guiglion et al. 2019, Minchev et al. 2019). Here, we propose a new explanation for the lithium ISM decline at high-metallicities, based on the importance of novae as producers of lithium.
\\In fact, novae are important sources of lithium in the Galaxy. In literature, D'Antona \& Matteucci (1991) first included novae into a detailed chemical evolution model: they considered as $^7$Li producers Asymptotic Giant Branch stars (AGB), classical novae and carbon stars, and concluded that novae could be important Li producers especially to explain the steep rise of Li abundance at [Fe/H]>-1.0 dex. A few years later, Romano et al. (1999) took into account also Li production from Type II SNe and Galactic Cosmic Rays (GCRs), and implemented detailed nova yields coming from 1D hydrodynamic models by Jose' \& Hernanz (1998) in the GCE model. They concluded that the most important $^7$Li producers were novae and GCRs (see also Romano et al. 2001; 2003). More recently, Matteucci (2010), Izzo et al. (2015) and Cescutti \& Molaro (2019) underlined again the importance of novae to explain lithium evolution.
\\The detections of $^7$Be (later decaying into $^7$Li) and $^7$Li in nova ejecta by Tajitsu et al. (2015) and Izzo et al. (2015), respectively, reinforced the idea that novae are important sources of lithium (see also Tajitsu et al. 2016; Molaro et al. 2016; Selvelli et al. 2018). These pieces of observational evidence are extremely important. On a theoretical side, in fact, it is well-known that Li can be produced in stars, either through the "Cameron-Fowler conveyor" (Cameron \& Fowler 1971), acting in intermediate-mass stars on the AGB (Sackmann \& Boothroyd 1992), or through the cool bottom process combined to some extra deep mixing in low-mass stars climbing the red giant branch (RGB; Sackmann \& Boothroyd 1999), or during thermonuclear runaways in nova outbursts (e.g. Starrfield et al. 1978), or through the $\nu$-process (first hypothesised by Domogatskii et al. 1978) in the He-shell of core-collapse supernovae (SNe). Moreover, a fraction (up to 20-30\%) of the meteoritic Li comes from spallation processes triggered by the energetic nuclei of GCRs in the ISM (Reeves et al. 1970; Meneguzzi et al. 1971; Lemoine et al. 1998; Romano et al. 2001; Prantzos 2012). The observational evidence for Li production, however, remains elusive. While it is known that a (small) fraction of RGB and AGB stars are Li-rich (e.g. Kirby et al. 2016, and references therein), the Li production from these stars can be hardly quantified, mostly because of the severe uncertainties on their mass loss rates (see, e.g., Romano et al. 2001; Travaglio et al. 2001, and discussions therein). As regards core-collapse SNe, to the best of our knowledge Li has never been found in their spectra. Thus, the detection of huge amounts of Li in the ejecta of classical novae (in excess of 1D theoretical model calculations) offer the only firm solution to the Galactic lithium problem, and point to novae as the main sources of Li in the Galaxy (Izzo et al. 2015, and discussion therein; see also Cescutti \& Molaro 2019).
\\The aim of this work is to study the chemical evolution of lithium by means of detailed chemical evolution models in the light of the newest observational data. In particular, we will focus on the decrease of lithium at high metallicities, which is still a topic of lively debate. Here, we test the hypothesis that the fraction of binary systems giving rise to novae is lower at higher metallicities, as suggested by the studies of Gao et al. (2014, 2017) and Yuan et al. (2015). A similar assumption of a metallicity-dependent occurrence probability has been tested by Simonetti et al. (2018) with respect to neutron star mergers.
\\The paper is organized as follows. In Section 2, we describe the observational data which have been considered to compare with the predictions of our chemical evolution models. In Section 3, we present the models adopted in this work. In Section 4, we show our results, that include some predictions to be tested by future observations. Finally, in Section 5, we draw our conclusions, based on the comparison between model predictions and available observational data.

\section{Observational data}

In this work, to study the Galactic lithium evolution we consider different datasets from literature. In particular, for the Galactic halo we consider the data of Charbonnel \& Primas (2005) and Sbordone et al. (2010).
\\Charbonnel \& Primas (2005) revised a large collection of Li measurements for halo stars from the literature, paying particular attention to the quality of the data and exploring in detail the temperature scale issue. NLTE corrections were applied to the Li abundances. Li determinations for stars in their "clean sample" (see Charbonnel \& Primas 2005, for details) are consistent with no dispersion on the plateau. We enlarge this dataset with VLT-UVES Li abundances for 28 halo dwarfs in the metallicity range -3.5 < [Fe/H] <-2.5 by Sbordone et al. (2010), that are measured by means of 3D hydrodynamical spectral synthesis including NLTE. A bending of the Spite plateau below [Fe/H] $\sim$ -3 is clearly present in the data of Sbordone et al. (2010).
\\For the discs, we adopt the recent spectroscopic data from the AMBRE Project (Guiglion et al. 2016) and GES (Fu et al. 2018), where a distinction is made between thick and thin disc stars, basing on chemical criteria. The AMBRE catalogue consists of Li abundances for 7300 stars, homogeneously derived from high-resolution spectra in the ESO archive with an automatic method (see Guiglion et al. 2016 for a description of the method and of the data validation). Overall, the Li abundance in the local ISM is found to increase from [Fe/H] = -1 to 0 dex, while it clearly decreases in the super-solar metallicity regime. The Li content of thin-disc stars displays a much steeper increase with [Fe/H] than that of thick-disc stars, that is found to increase only slightly with time (metallicity). These results are confirmed by the analysis of Fu et al. (2018), which discuss Li measurements for main-sequence field stars from the GES iDR4 catalogue pointing out the higher level of Li enrichment in thin-disc stars, with respect to the thick-disc population. These authors also mention the possibility that the decrease of Li abundance at super-solar metallicities can be due to a reduced nova rate at high [Fe/H], coupled to Li depletion -rather than production- in AGB stars (on the latter point, see also Romano et al. 2001), a possibility that we fully address here with our GCE models.
\\For the meteoritic value, we use the determination by Lodders et al. (2009).
\\For the Galactic bulge we adopt the data of Gonzalez et al. (2009) and Bensby et al. (2011). However, the 13 stars with measured Li abundances in the sample of Gonzalez et al. (2009) are all RGB/AGB stars in which the initial atmospheric Li content has been altered, so they can not be used to safely constrain our evolutionary models for the bulge. The sample analysed in Bensby et al. (2011), consisting of 26 microlensed dwarf and subgiant stars, offers, in principle, a better hope. 1D, NLTE Li abundances (based on MARCS models) could be derived, however, only for 5 objects. Out of these, only 2 have T$_{eff}$ $>$ 5900 K, thus further reducing the number of stars that can be taken as reliable tracers of bulge Li enrichment (see Bensby et al. 2011, and references therein). We discuss our predictions about the evolution of Li in the Galactic bulge in Section 4; we caution that these still wait for a proper dataset in order to be confirmed or disproved.

\section{Models}

In this work, we adopt the following chemical evolution models for the Galactic halo, discs and bulge.
\begin{itemize}
\item Two-infall model (Chiappini et al. 1997 and Romano et al. 2010). It assumes that the Milky Way forms by means of two major gas infall episodes: the first infall episode gave rise to the halo-thick disc, whereas during the second one, which is slower and delayed with respect to the first one, the thin disc forms.
\item Parallel model (Grisoni et al. 2017, 2018). It assumes that the thick and the thin disc stars formed out of two separate infall episodes in two distinct evolutionary phases, which evolve independently. This model was tested in Grisoni et al. (2017) for the solar neighborhood, and then extended to the other Galactocentric distances in Grisoni et al. (2018).
\item For the Galactic bulge, we consider the model by Matteucci et al. (2019) which assumes a fast formation on a short timescale and with high star formation efficiency ($\nu$=25 Gyr$^{-1}$, see Table 1).
\end{itemize}

\subsection{Model equations}

\begin{table*}
\caption{Input parameters for the best chemical evolution models. In the first column, there is the name of the model. In the second column, we show the adopted initial mass function. In the third column, there is  the star formation efficiency ($\nu$). In the fourth column, we give the timescales for mass accretion ($\tau$).}
\label{tab_01}
\begin{center}
\begin{tabular}{c|cccccccccc}
  \hline
%\noalign{\smallskip}
\\
 Model & IMF &$\nu$& $\tau$\\
&  &[Gyr$^{-1}$]& [Gyr]\\
\\
%\noalign{\smallskip}
\hline
%\noalign{\smallskip}

Two-infall & Kroupa et al. (1993) & 2 (halo-thick) & 1 (halo-thick) \\
	   & 			  & 1 (thin disc) & 7 (thin disc)  \\
%\noalign{\smallskip}
 \hline
%\noalign{\smallskip}

Thick disc & Kroupa et al. (1993) & 2& 0.5\\

%\noalign{\smallskip}
 \hline
%\noalign{\smallskip}

Thin disc & Kroupa et al. (1993) & 1  & 7 \\

%\noalign{\smallskip}
 \hline
%\noalign{\smallskip}

Bulge & Salpeter (1955) & 25  &0.1 \\

%\noalign{\smallskip}
 \hline
%\noalign{\smallskip}

\end{tabular}
\end{center}
\end{table*}

\begin{figure*}
\includegraphics[scale=0.39]{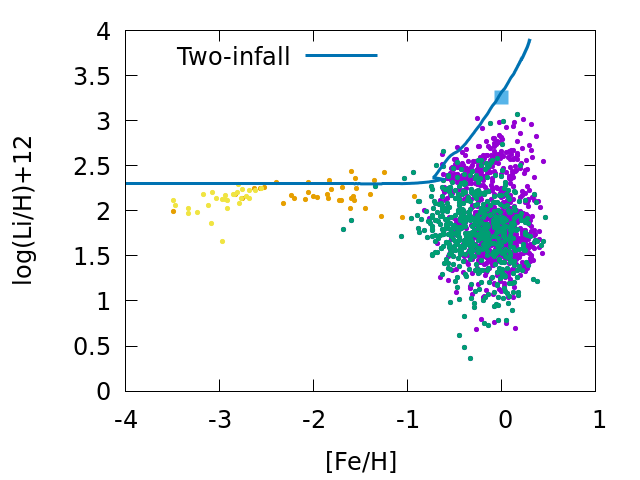}
\includegraphics[scale=0.39]{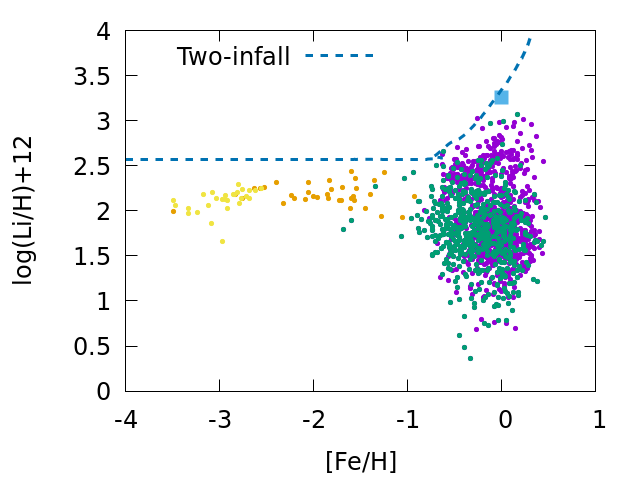}
 \caption{\textit{Left panel:} Observed and predicted lithium abundance as a function of metallicity in the solar neighbourhood. The predictions are from the two-infall model (blue line). The data are from Charbonnel \& Primas (2005) (orange dots), Sbordone et al. (2010) (yellow dots) and from GES (Fu et al. 2018) for the thick disc (green dots) and thin disc (purple dots). The meteoritic value is the one by Lodders et al. (2009) (light-blue square). \textit{Right panel:} Same as the left panel, when a high-primordial Li abundance is adopted.}
 \label{fig_01}
\end{figure*}

The time evolution of $G_i$, which is the mass fraction of the element $i$ in the gas, is described by the following equations (for a detailed description, we refer the reader to Matteucci 2012):
\begin{align} \label{eq_01}
\begin{split}
& \dot G_i(r,t)= -\text{SFR}(r,t) X_i(r,t)+R_i(r,t)+ \dot G_i(r,t)_{inf}
\end{split}
\end{align}
where $\text{SFR}(r,t)$ corresponds to the star formation rate (SFR), $X_i(r,t)$ indicates the abundance by mass of the element $i$, $R_i(r,t)$ represents the rate of matter restitution from stars with different masses into the ISM, and $\dot G_i(r,t)_{inf}$ gives the rate of gas infall.
\\The SFR is given by the Schmidt-Kennicutt law (Kennicutt 1998):
\begin{equation} \label{eq_03}
\text{SFR}(r,t)=\nu \sigma_{gas}^k(r,t),
\end{equation}
where $\sigma_{gas}$ indicates the surface gas density, $k=1.4$ represents the exponent of the law and $\nu$ corresponds to the star formation efficiency (SFE), which is fixed in order to reproduce the SFR at present time. For the initial mass function (IMF), we use the Kroupa et al. (1993) one for the halo, thick and thin discs, and the Salpeter (1955) one for the bulge.
\\In the term $R_i(r,t)$ of Eq.\,\eqref{eq_01}, we account for detailed nucleosynthesis from low and intermediate mass stars, super-AGB stars, Type Ia SNe (which originate from white dwarfs in binary systems) and Type Ib, Ic, II SNe (which originate from core-collapse massive stars).
\\In this work, we focus also on the contribution of novae, which are important lithium producers. Novae are binary systems of a white dwarf (WD) and a low mass main sequence star. The novae rate is computed by assuming that it is proportional to the formation rate of C-O WDs (see Matteucci 2012, Spitoni et al. 2018). In particular:
\begin{equation} \label{eq_novae}
R_{\text{nova}}(t)=\alpha \int\limits^{8}_{0.8}\text{SFR}(t-\tau_{m2}-\Delta t)\phi(m)dm,
\end{equation}
where $\alpha$ is the fraction of WDs in binary systems giving rise to novae, $\tau_m$ is the lifetime of WD progenitors of mass $m$, $\Delta t$ = 1 Gyr is a suitable average cooling time (see Romano et al 1999, and references therein) and $\phi$ is the stellar IMF. Following Bath \& Shaviv (1978), each nova is supposed to suffer 10$^4$ eruptions during its life. For the sake of simplicity, we consider all the outbursts to happen at the same time, i.e. at the time of the formation of the nova system (this means no time delay between successive outbursts, but see Cescutti \& Molaro 2019).
\\Finally, the last term in Eq.\,\eqref{eq_01} corresponds to the gas infall rate. In the case of the two-infall model, the gas infall law is as follows:
\begin{align} \label{eq_02}
\dot G_i(r,t)_{inf}=A(r)(X_i)_{inf}e^{-\frac{t}{\tau_1}}+B(r)(X_i)_{inf}e^{-\frac{t-t_{max}}{\tau_2}},
\end{align}
where $G_i(r,t)_{inf}$ refers to the infalling material in the form of element $i$ and $(X_i)_{inf}$ represents the composition of the infalling gas, which we assume to be primordial. The parameter t$_{max}$ indicates the delay of the beginning of the second infall episode. The parameters $\tau_1$ and $\tau_2$ are the timescales of gas accretion for the halo-thick and thin discs, respectively. The timescales of gas accretion are free parameters in the model, and they have been tuned in order to fit the observed metallicity distribution function in the solar vicinity (see Table 1). The coefficients $A(r)$ and $B(r)$ are chosen in order to reproduce the total surface mass density at present time in the solar neighborhood, and we follow the prescriptions of Romano et al. (2000).
\\On the other hand, in the case of the parallel model, since we assume two separate infall episodes, the gas infall law is given by:
\begin{equation} \label{eq_07}
(\dot G_i(r,t)_{inf})|_{thick}=A(r)(X_i)_{inf}e^{-\frac{t}{\tau_1}},
\end{equation}
for the thick disc and
\begin{equation} \label{eq_08}
(\dot G_i(r,t)_{inf})|_{thin}=B(r)(X_i)_{inf}e^{-\frac{t}{\tau_2}},
\end{equation}
for the thin disc, respectively. The coefficients $A(r)$ and $B(r)$ and the timescales $\tau_1$ and $\tau_2$ have the same meaning as explained for Eq. \,\eqref{eq_02}, and we follow the prescriptions of Grisoni et al. (2017; 2018), Matteucci et al. (2018).

\subsection{Nucleosynthesis prescriptions}

\begin{figure*}
\includegraphics[scale=0.3]{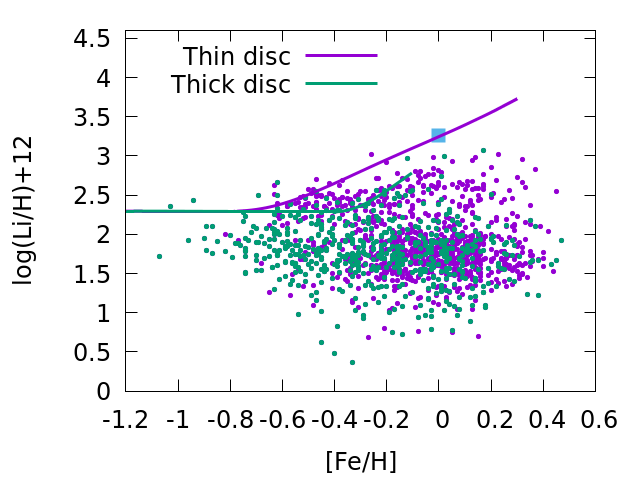}
\includegraphics[scale=0.3]{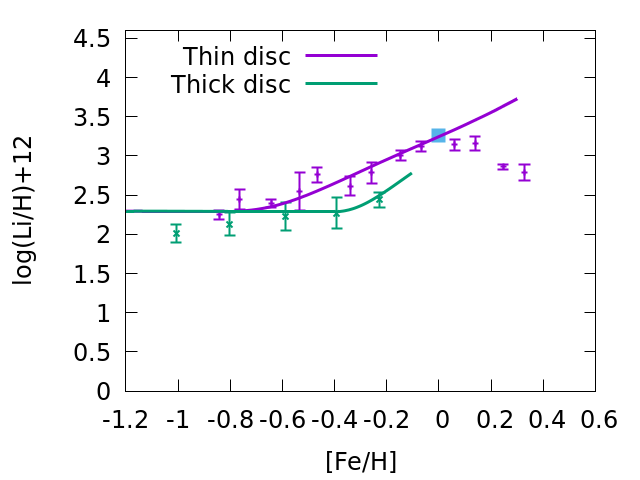}
\includegraphics[scale=0.3]{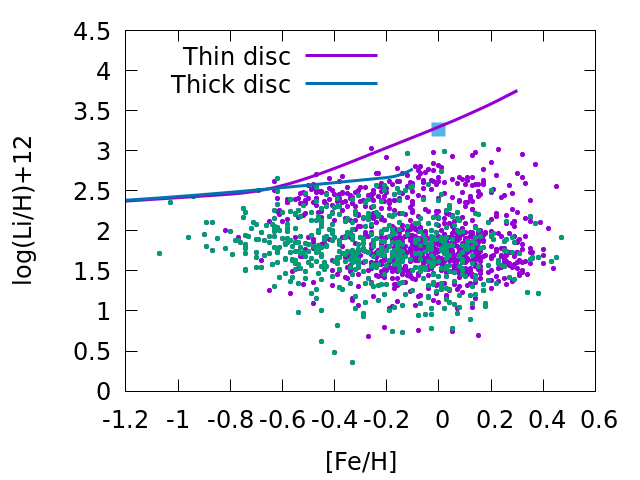}
\includegraphics[scale=0.3]{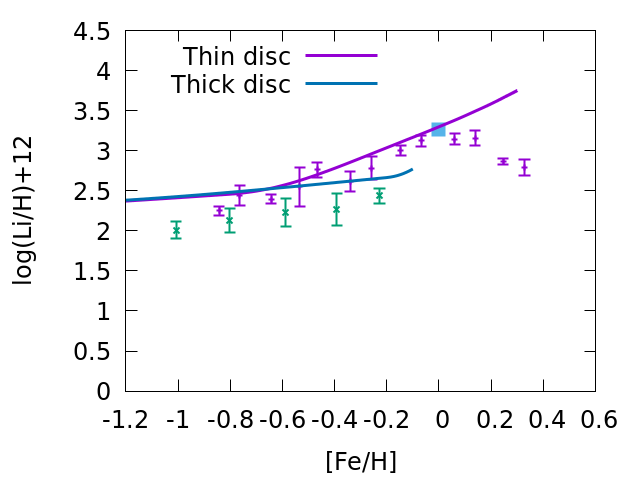}
 \caption{\textit{Left-hand panels:} Observed and predicted lithium abundance as a function of metallicity for thick- and thin-disc stars in the solar neighbourhood. The predictions are from the parallel model for the thick (green line) and thin disc (purple line). The data are from GES (Fu et al. 2018) for the thick disc stars (green dots) and thin disc stars (purple dots). The meteoritic value is the one by Lodders et al. (2009) (light-blue square). \textit{Right-hand panels:} same as the left panel, but with AMBRE data (Guiglion et al. 2016) for the thick disc stars (green dots) and thin disc stars (purple dots).}
 \label{fig_02}
\end{figure*}

Here, we adopt the following Li nucleosynthesis sources and prescriptions.
\begin{itemize}
\item Ventura et al. (2013 + private communication) for LIMS (1-6 MSun) and super-AGB (6-8 MSun);
\item Nomoto et al. (2013) for core-collapse SNe. This contribution is very uncertain and, moreover, it turns out to be a minor one; therefore, it is suppressed in our best models;
\item The Li ejected during one nova outburst (the total number of outbursts in a nova life is 10$^4$) is assumed to be in the range given by Izzo et al. (2015), where it was measured that M$_{Li}$=0.3-4.8 10$^{-10}$ M$_{\bigodot}$ in the ejecta of nova V1369 Cen.
\item GCR as in Smiljanic et al. (2009), where it was derived the relation for $^9$Be:
\begin{equation} \label{eq_09}
\text{log(Be/H)=-10.38+1.24[Fe/H]}.
\end{equation}
Then assuming a scaling ratio of $^7$Li/$^9$Be$\sim$7.6 (see Molaro et al. 1997), it is possible to get the relation also for $^7$Li (see also Cescutti \& Molaro 2019):
\begin{equation} \label{eq_09}
\text{log(Li/H)=-9.50+1.24[Fe/H]}.
\end{equation}
\end{itemize}

\section{Results}

In this section, we show the results based on the comparison between model predictions and observations for the various Galactic components: halo, thick and thin discs, and bulge. In Table 1, the input parameters of the different models are listed. In the first column, there is the name of the model. Then, we indicate the adopted IMF, the star formation efficiency ($\nu$) and the timescale of formation ($\tau$) of the Galactic components.

\begin{figure*}
\includegraphics[scale=0.39]{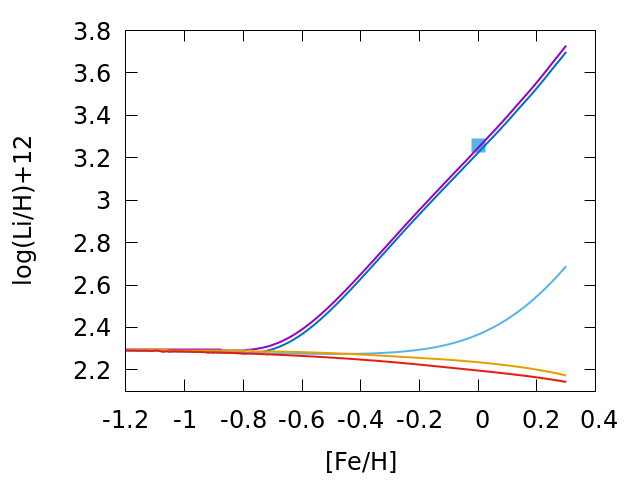}
 \caption{Predicted lithium abundance as a function of metallicity for the thin disc, where the various lithium sources are isolated. The predictions are from the parallel model for the thin disc with all the sources (purple line), only novae (blue line), only GCR (light blue line), only AGB (orange line) and only astration (red line). The meteoritic value is the one by Lodders et al. (2009) (light-blue square).}
 \label{fig_03}
\end{figure*}

\subsection{The Galactic halo}

In Fig. \ref{fig_01}, we show the observed and predicted lithium abundance as a function of [Fe/H] as index of metallicity for solar neighbourhood stars. We consider the data from Charbonnel \& Primas (2005) (orange dots), Sbordone et al. (2010) (yellow dots) and from GES (Fu et al. 2018) for the thick disc (green dots) and thin disc (purple dots). The predictions are from the two-infall model, when a primordial lithium abundance of A(Li)$\sim$2.3 (left panel) or A(Li)$\sim$2.6 (right panel) is adopted, respectively.
\\By assuming a primordial Li abundance of A(Li)$\sim$2.3, and Li production from LIMS, super-AGB stars, novae and GCRs as described in Sect. 3.2, we can reproduce the upper envelope of the observational data, as well as the meteoritic value (that is immune from destruction processes) with our model. Still, we must invoke some internal destruction mechanism(s) in the most metal-poor stars to explain the bending of Li abundance at the low metallicity end. If we consider a primordial Li abundance of A(Li)$\sim$2.6 as suggested by SBBN and the measurements of Planck and WMAP, instead, the predicted plateau requires the activation of Li destruction channels in all metal-poor halo dwarfs to be made consistent with the data (see Fu et al. 2015).
\\Because of Li depletion acting in subsequent generations of stars during the whole Galactic evolution, the rise from the plateau requires the same contributors to the Li synthesis, independently of the assumed primordial value of A(Li) (see the discussion in Romano et al. 2003). In particular, the rise from the primordial value is always explained as due to the fundamental contribution of long-lived stellar sources (Romano et al. 1999, 2001). Thus, from now on we do not focus anymore on the discrepancy between the primordial Li values, but concentrate on the Galactic Li evolution. In particular, we set the primordial Li abundance to A(Li)$\sim$2.3 and focus on the different Li producers, remembering that the conclusions on the Galactic lithium evolution will not be affected by our particular choice of the primordial Li abundance.
\\A characteristic feature of the two-infall model that we can appreciate in Fig. \ref{fig_01} is the back and forth loop at [Fe/H]$\sim$-0.8 dex. This is related to the transition between the halo-thick disc and the thin disc phases. When the formation of the inner halo and thick disc is terminated, in fact, the star formation stops. A new episode of gas infall starts at this point, which provides large amounts of fresh, unprocessed gas. As a consequence, the metallicity of the ISM first suddenly decreases, then, as soon as the star formation is reactivated, it increases again. This happens about 1 Gyr after the beginning of the Galaxy formation, that is also the time at which novae start to contribute to the Galactic Li enrichment in our model (see Eq. 3). Therefore, a rapid increase of the ISM Li abundance is predicted to start with the formation of the thin disc component. 
\\In the thin disc phase, there is the rise which is mainly due to novae (Romano et al. 1999, 2001; Izzo et al. 2015; Cescutti \& Molaro 2019). In this way, the meteoritic value (Lodders et al. 2009) can be reached. We notice that after the meteoritic value is reached, the model predictions still rise at variance with the observations that show a decline (GES data of Fu et al. 2018, but the decline is evident also in the data of Guiglion et al. 2016, Buder et al. 2018, Bensby \& Lind 2018). To better study what happens in the Galactic discs, we apply the parallel approach of Grisoni et al. (2017; 2018) (see next section).

\subsection{The Galactic disc(s)}

\begin{figure*}
\includegraphics[scale=0.39]{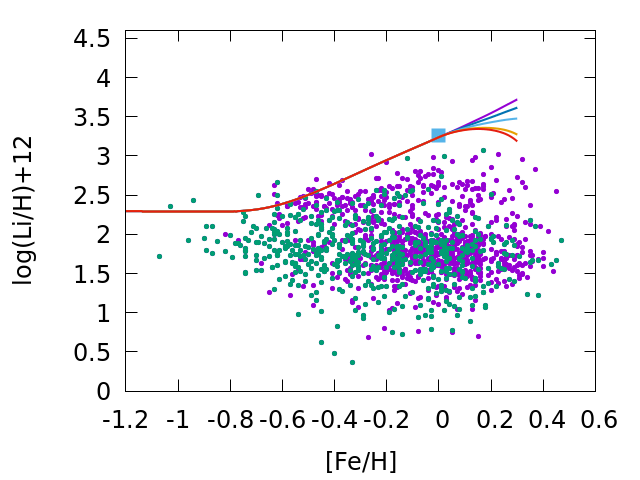}
\includegraphics[scale=0.39]{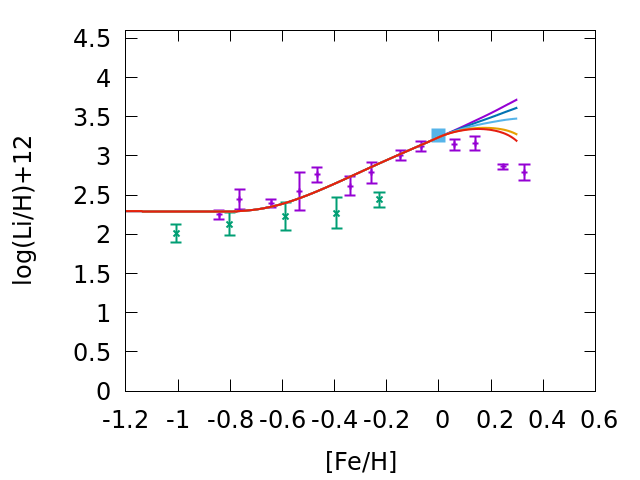}
 \caption{\textit{Left panel}: Observed and predicted lithium abundance as a function of metallicity for the thin disc in the case with variable fraction of binary systems giving rise to novae. The predictions are from the parallel model for the thin disc with constant $\alpha$ (purple line) and with different variable $\alpha$ laws. The data are from GES (Fu et al. 2018) for the thick disc stars (green dots) and thin disc stars (purple dots). The meteoritic value is the one by Lodders et al. (2009) (light-blue square). \textit{Right panel}: same as the left panel, but with AMBRE data (Guiglion et al. 2016) for the thick disc stars (green dots) and thin disc stars (purple dots).}
 \label{fig_04}
\end{figure*}

In Fig. \ref{fig_02}, we show the predicted and observed lithium abundance as a function of metallicity for the thick and thin discs. The predictions are from the parallel model (Grisoni et al. 2017, 2018). This model was tuned to reproduce the [Mg/Fe] vs. [Fe/H] for the thick and thin disc stars observed by AMBRE (Mikolaitis et al. 2017). In fact, in the [Mg/Fe] vs. [Fe/H] diagram it is possible to see two distinct sequences corresponding to the thick and thin disc stars, with the thick disc been $\alpha$-enhanced due to a shorter timescale of formation and higher star formation efficiency with respect to the thin disc. These two sequences corresponding to thick and thin discs stars have been observed also by other surveys, such as APOGEE (Hayden et al. 2015) and GES (Rojas-Arriagada et al. 2017). A dichotomy between thick and thin discs is observed also in the A(Li) vs. [Fe/H] plane, as clearly shown by the data of Fu et al. (2018) and Guiglion et al. (2016) (Fig. \ref{fig_02}, left-hand and right-hand panels, respectively) and has been studied by Prantzos et al. (2017) and Cescutti \& Molaro (2019) by means of chemical evolution models.
\\Here, we consider the models of Grisoni et al. (2017, 2018) to explore this dichotomy between the thick and thin discs. In Fig. \ref{fig_02}, we show our results. Since the evolution of the thin disc happens on relatively long timescales, novae can contribute to lithium enrichment in this Galactic component much more than they do in the thick disc, that is evolving at a quicker pace (see Table 1). With these models, we can reproduce the plateau at low metallicities, as well as the subsequent rise. In the lower panels of Fig. \ref{fig_02}, we show the case when the contribution from massive stars is taken into account, while this contribution is suppressed in the model predictions displayed in the upper panels. We can see that for the thin disc the results remain almost unchanged, because of the overwhelming production from novae and GCRs. For the thick disc, instead, we get a mild enhancement since in this case we have a faster evolution and, hence, basically no contribution to Li enrichment from novae, but a higher number of contributing massive stars. In the following, when talking about the thin disc we will not take into account the contribution by massive stars since the results for this Galactic component remain almost unchanged (see also model B of Prantzos et al. 2012, as well as Cescutti \& Molaro 2019 that do not take into account the contribution from massive stars in their models).
\\In Fig. \ref{fig_03}, we show the different contributions from the different lithium sources: all sources, novae, GCR and AGB/super-AGB. The figure is similar to Fig. 1 of Cescutti \& Molaro (2019), and also in our case we can see that the main contribution comes from novae which are the main responsible for the rise at [Fe/H]$\sim$-0.6 dex. In the following, we summarize the contribution from the various sources to the meteoritic lithium content. About 10\% of the meteoritic $^7$Li comes from primordial nucleosynthesis, after taking into account all factors affecting the evolution (and therefore also astration). Then, 16\% is due to GCR. Therefore, more then 70\% comes from stellar sources (in agreement with Prantzos 2012). In particular, novae represent our main source of lithium. Overall, from the analysis of the various lithium producers, we confirm that novae are a fundamental source of lithium in the Galaxy, in agreement with other previous studies (Romano et al. 1999, 2001, 2003; Matteucci 2010). In particular, they are the most important one (Izzo et al. 2015 and then also Cescutti \& Molaro 2019).
\\The models discussed up to now can reproduce the plateau at low metallicity, as well as the rise at disc metallicities. However, they cannot reproduce the decrease at super-solar metallicities which, as recalled in the Introduction, has been claimed by recent observations, such as GES (Fu et al. 2018) and AMBRE (Guiglion et al. 2016). To reproduce this decrease, we assume that the fraction of binary systems giving rise to novae is no more constant, but it decreases at high metallicities, as suggested by the studies of Gao et al. (2014, 2017) and Yuan et al. (2015).
\\In Fig. \ref{fig_04}, we show the predicted and observed lithium abundance as a function of metallicity, under different laws for the fraction of binary systems giving rise to novae, namely $\alpha$ (see Eq. 3). In the case with constant $\alpha$, we considered M$_{Li}$=4.8 10$^{-10}$ M$_{\bigodot}$ (maximum value from Izzo et al. 2015) and $\alpha$=0.017 to reproduce the present time novae rate (R$_{\text{novae}} \sim 20-30$ yr$^{-1}$; see also Izzo et al. 2015). Then, we consider variable $\alpha$ laws. In particular, we take for the mass produced by each nova M$_{Li}$=0.8 10$^{-10}$ M$_{\bigodot}$ (also in the range measured by Izzo et al. 2015) and we assume that:

\begin{equation} \label{eq_alfa}
\alpha=\begin{cases} 0.1, & \mbox{if }\mbox{[Fe/H]} \le 0 \\ 0.1-\beta\mbox{[Fe/H]}, & \mbox{if }\mbox{[Fe/H]}>0
\end{cases} 
\end{equation}

\begin{figure*}
\includegraphics[scale=0.39]{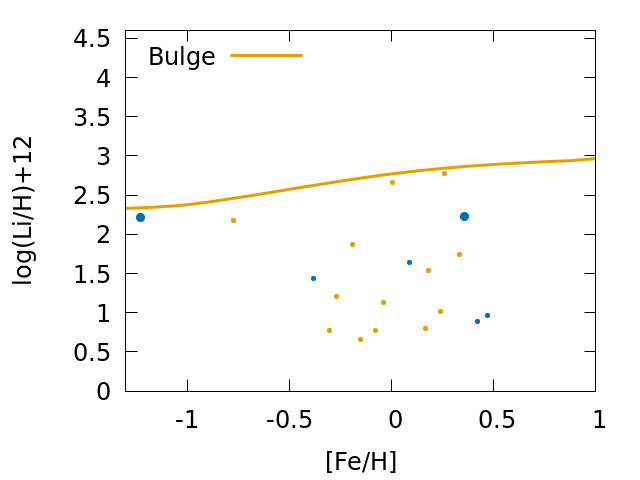}
 \caption{Observed and predicted lithium abundance as a function of metallicity in the Galactic bulge. The predictions are from the model of the Galactic bulge of Matteucci et al. (2019). The data of bulge stars are from Gonzalez et al. (2009) (orange dots) and Bensby et al. (2011) (blue dots, with the two bigger dots corresponding to the two stars which have T$_{eff}$ $>$ 5900 K).}
 \label{fig_06}
\end{figure*}

where $\alpha$(>0) is constant up to [Fe/H]=0 and then decreases linearly with metallicity. We test different $\beta$ values ($\beta$=0.1, 0.2, 0.3, 0.33) and therefore different laws, as it can be seen in the figure. With this assumption, we get the plateau, the rise, and then also the bending which is due to the lower fraction of binary systems at the higher metallicities according to Eq. 9. We stress that, given the uncertainties involved in this kind of calculations, our results are qualitative, rather than quantitative. A better understanding of the formation and evolution of the nova systems is needed before we can work out a more refined model.  
\\Other possible explanations that have been proposed for the decrease at high metallicities are: i) lower yields of Li from single stars at high metallicity (Prantzos et al. 2017), even if no physical justification for this fact can be found, and ii) the effect of radial migration of stars reflecting different evolutionary paths from the inner thin disc regions (Guiglion et al. 2019).
\\Some uncertainties might be introduced by the Li depletion in the super-solar metallicity stars. Li depletion during the stellar evolution is the link between the ISM Li abundance and the current Li abundance measured in dwarf stars. It is expected to be small in principle. The two processes that are responsible for reducing the surface chemical abundances are microscopic diffusion and nuclear burning. The first mechanism takes the surface element to the base of the surface convective zone in a long timescale and is weakened with strong mixing.  Fu et al (2018) show that main sequence diffusion in a 7-10 Gyr time reduces the surface Li abundance of a low mass solar metallicity star for $\lesssim$ 0.2 dex. The latter process, with the reaction $^7Li(p,\gamma)^4He+^4He$, burns Li in the surface convective zone. Metal-poor and Solar metallicity main sequence stars (M$\gtrsim 0.7 M_\odot$) have a very thin surface convective zone, even the bottom of the convective zone is not hot enough to burn Li efficiently. In contrast, super-solar metallicity stars have a relatively extended surface convective zone even during the main sequence and will eventually burn some Li. Unfortunately, standard stellar models are not applicable for Li evolution at super-solar metallicity though safe for the more metal-poor stars and most of the other elements. Standard stellar models homogenize the convective zone and treat it as a single radiative zone, thus the surface Li is erased almost immediately because of the high temperature at the base of the deep convective zone. In reality, the time scale of the nuclear reactions is comparable to the mixing time scale, Li burns in a mild temperature gradient during the convection. To quantitatively model Li evolution at this metallicity, new stellar models with the so-called "diffusive convection" is needed. Before the recent large Galactic surveys we discussed before, super-solar metallicity stars are rarely studied in the literature. Now it is the golden time to call for stellar models optimized for super-metallicity, and the new stellar modeling results will help to examine the nova rate law we use (Eq. 7) in the Galactic chemical evolution models.

\subsection{The Galactic bulge}

Finally, we show also the results in the case of the Galactic bulge.
\\The model for the Galactic bulge used here is the one by Matteucci et al. (2019), which assumes a fast formation on a short timescale and with high star formation efficiency ($\tau$=0.1 Gyr$^{-1}$ and $\nu$=25 Gyr$^{-1}$), and a flatter IMF with respect to the solar-vicinity, i.e. a Salpeter (1955) IMF instead of the Kroupa et al. (1993) one (see Table 1). This model reproduces the [Mg/Fe] vs. [Fe/H] relation, as well as the metallicity distribution function (MDF) of the bulge stars observed by GES (Rojas-Arriagada et al. 2017) and APOGEE (Rojas-Arriagada et al. 2019), and it refers to the real bulge stars, i.e. the so-called "classical" bulge. In fact, in the Galactic bulge, there is the possibility of a different stellar population originating via secular evolution from the inner disc and cohabiting with the bulge stars formed in situ, although firm conclusions are still not reached. Here, we take into account only the classical bulge population, which should be the dominant one.
\\In Fig. \ref{fig_06}, we show the predicted A(Li) vs. [Fe/H] relation in the bulge in comparison to observations of bulge stars. In particular, we adopt the data of Gonzalez et al. (2009) (orange dots) and Bensby et al. (2011) (blue dots, with the two bigger dots corresponding to the two stars which have T$_{eff}$ $>$ 5900 K). As regards to the model predictions, the Galactic bulge has evolved much faster than the disc and long-lived stars (in particular novae), which are fundamental lithium producers, did not have time to enrich the ISM in lithium. The mild Li enrichment that we predict comes from the contributions of massive stars. Our results are in agreement with the ones by Grieco et al. (2012), which performed a similar theoretical study of lithium in the Galactic bulge. However, we do not exclude the possibility to find some lithium-rich stars also in the Galactic bulge, which could have come via secular evolution from the inner disc. In fact, Matteucci et al. (2019) has concluded that there can be a different stellar population originating via secular evolution from the inner disc and cohabiting with the bulge stars formed in situ.
\\We can see that up to now the agreement between the classical bulge model and the bulge data is rather good, but more data will be necessary to draw firm conclusions about the evolution of lithium in the Galactic bulge.

\section{Conclusions}

In this work, we have studied the evolution of lithium in the Milky Way halo, discs and bulge. In particular, we have focused on the puzzling decrease of lithium at high metallicity. We considered the most recent observational data from Galactic stellar surveys and we compared the observations with our detailed chemical evolution models. The adopted models have been already tested on the evolution of the $\alpha$-elements and Fe in the thick and thin discs (Grisoni et al. 2017, 2018), as well as the bulge (Matteucci et al. 2019).
\\Our main results can be summarized as follows.
\begin{itemize}
\item We confirm that novae are important sources of lithium, as pointed out by previous studies (D'Antona and Matteucci 1991; Romano et al. 1999, 2001, 2003; Matteucci 2010). In particular, they are the most important source, in agreement with the recent results by Izzo et al. (2015, see their figure 5) and then also Cescutti \& Molaro (2019). These conclusions are supported by the Be and Li line identifications in nova ejecta by Tajitsu et al. (2015) and Izzo et al. (2015), respectively, that reinforced the idea that novae are fundamental sources of lithium.
\item Concerning the decrease of Li at high metallicities in the thin disc, we propose a novel explanation. In particular, we show that this can be due to a lower fraction of binary systems giving rise to novae at high metallicities. This assumption of a metallicity dependent occurrence probability for this kind of systems is supported observationally by the studies of Gao et al. (2014, 2017) and Yuan et al. (2015), and it can be crucial to explain the decrease of lithium at high metallicities.
\\Other alternative explanations that have been proposed in the literature to explain this feature by means of chemical evolution models are: i) lower yields of Li from stars at high metallicities, even if no physical reasons for this fact can be found (Prantzos et al. 2017); and ii) stellar migration of stars coming from the inner regions of the Milky Way disc (Guiglion et al. 2019).
\item We also considered the lithium evolution in the Galactic bulge. In particular, we consider a model for the "classical" bulge as in Matteucci et al. (2019) with very high SFR compared to the other components, and we showed that the Galactic bulge has evolved much faster and long-lived stars (in particular novae), which are fundamental lithium producers, did not have time to enrich the ISM in lithium. The mild Li enrichment that we observe could have come from the contributions of massive stars, since long-lived stars did not have had time to contribute. However, we do not exclude the possibility to find some lithium-rich stars also in the Galactic bulge, which could have come via secular evolution from the inner disc. However, we still need further data to draw firm conclusions about lithium evolution in this Galactic component.
\end{itemize}

\section*{Acknowledgments}

V.G. and F.M. acknowledge financial support from the University of Trieste (FRA2016).
\\We are thankful to the anonymous referee for carefully reading our manuscript.

\end{document}